\documentclass[useAMS,usenatbib]{mn2e}
 \usepackage{graphicx}
\title[Detection of CO absorption in the atmosphere of the hot Jupiter HD~189733b]{Detection of CO absorption in the atmosphere of the hot Jupiter HD~189733b}
\author[Rodler et al.]{F. Rodler$^{1}$\thanks{E-mail:
rodler@ieec.uab.es}, M. K\"urster$^{2}$ and J.~R. Barnes$^{3}$\\
$^{1}$Institut de Ci\`encies de l'Espai (CSIC-IEEC), Campus UAB, Facultat de Ci\`encies, Torre C5 parell, 2$^{a}$ planta, 08193, Bellaterra, Spain\\
$^{2}$Max-Planck Institut f\"ur Astronomie, K\"onigstuhl 17, D-69117 Heidelberg, Germany\\
$^{3}$Centre for Astrophysics Research, University of Hertfordshire, College Lane, Hatfield (Herts.) AL10 9AB, UK}
\begin{document}

\date{Accepted 1988 December 15. Received 1988 December 14; in original form 1988 October 11}

\pagerange{\pageref{firstpage}--\pageref{lastpage}} \pubyear{2013}

\maketitle

\label{firstpage}

\begin{abstract}
With time-series spectroscopic observations taken with the Near Infrared Spectrometer (NIRSPEC) at Keck II, we investigated the atmosphere of the close orbiting transiting extrasolar giant planet, HD~189733b. In particular, we intended to measure the dense absorption line forest around 2.3~micron, which is produced by carbon monoxide (CO). CO is expected to be present in the planetary atmosphere, although no detection of this molecule has been claimed yet. To identify the best suited data analysis method, we
 created artificial spectra of planetary atmospheres 
and analyzed them by three approaches found in the literature, the deconvolution method, data modeling via $\chi^2$-minimization, and cross-correlation. As a result, we found that cross-correlation and $\chi^2$-data modeling show systematically a higher sensitivity than the deconvolution method. We analyzed the NIRSPEC data with cross-correlation and detect CO absorption in the  day-side spectrum of HD~189733b at the known planetary radial velocity semi-amplitude with  $3.4\sigma$ confidence.    
\end{abstract}


\section{Introduction}

The study of exoplanets is one of the most vibrant and exciting fields in modern astronomy.
In the past 18 years, more than 860 exoplanets have been discovered\footnote{website: http://www.exoplanet.eu (2013-02-27)}. This has led
to an increasing interest in the physical characterization of these new objects. 
As result of those characterization efforts, several chemicals have thus far been detected in
the atmospheres of a few planets, as shown below. 

Most of the planets discovered so far are located too close to their host stars to appear separated. Consequently, the light coming from the planet and star are observed simultaneously. When attempting to measure the light only from the planet, one has to find a strategy to remove the dominating star light. A prosperous strategy to learn more about the atmospheres of remote planets is through investigation of those planets that transit their parent stars.
The atmospheric properties of transiting exoplanets can be measured in two ways: (1) during the passage of the planet in front of the star (transit), when part of the stellar light crosses the planetary atmosphere and the signature of chemicals in that atmosphere gets imprinted in the light we measure from the star, and (2) during the passage of the planet behind the star (eclipse), when the star temporarily blocks the planet's emission and we can determine its temperature, albedo, chemical composition, etc., from the difference spectrum. Via transits, NaI \citep{Char02}, HI, OI and CII \citep{Vida03,Vida04}, water vapor \citep{07} have been detected from space in the atmospheres of the hot Jupiters HD~209358b and HD~189733b. From the ground, \citet{Redf08} measured NaI in the atmosphere of HD~189733b, while \citet{Snel08,Sne10} confirmed the Charbonneau et al. (2002) detection of NaI in HD~209458b and also detected CO via  the analysis of the transmission spectrum using high-resolution spectroscopy around 2.3~$\mu$m. The latter result allowed these authors to directly measure the RV of a transiting hot Jupiter for the very first time. 
\citet{Bean10} investigated the atmosphere of the Super Earth GJ~1214b
 via transmission spectroscopy and found no evidence of atmospheric features, indicating a hydrogen atmosphere with high clouds, or a water dominated atmosphere \citep{deMo12}. Most recently, \citet{Sing11} and \citet{Colo12} announced the first detections of potassium in two planets, XO-2b and HD~80606b. Furthermore, eclipses have already provided temperature measurements for over twenty planets, both from space and more recently also from the ground (e.g. \citealp{Sing09,Lope10}).

In this work, we focus on a further strategy to measure atmospheric features in planets that is based on intermediate and high spectral resolution (i.e. $R=\lambda / \Delta \lambda > 20,000$; $\lambda$ denotes the wavelength) spectroscopy with very large telescopes. Key to this method is to observe a large number of spectral features coming from the planet and to observe them at different orbital phases so that the traveling faint planetary signal can be disentangled from the dominating stellar one. Considering this, the main advantage of this method is that it is not restricted to transiting planets (c.f. \citealp{Brog12,Rodl12b}).

In the past, this method was used in the optical with the goal to detect starlight reflected from hot Jupiters (i.e. massive planets that are a few stellar radii away from their host stars) and to measure their exact masses. Although all those campaigns resulted in a non-detection of reflected starlight, stringent upper limits to the planet-to-star flux ratio 
and to the geometric albedo of these planets could be established. To date, the tightest $3\sigma$-upper limits on the
geometric albedos of the hot Jupiters $\tau$~Boo~b, HD~75289b and $\upsilon$~And~b 
 are $0.39$ (Leigh et al.~2003a), $0.46$ (Rodler et al.~2008), and $0.42$ (Collier Cameron et al.~2002), respectively.
These results consequently provided
important constraints on models of the planetary atmospheres such as those by Marley et al.~(1999) and Sudarsky et al.~(2000, 2003). As a result,
 models that
predicted a high reflectivity for the planetary atmosphere could
be ruled out for the studied planets.

Towards near-infrared (NIR; $1-2.5~\mu$m) wavelengths, the planet-to-star flux ratios drastically increase due to the strong thermal emission of hot Jupiters. \cite{2001ApJ...546.1068W,Demi05,2007MNRAS.379.1097B,barn07b,bar08,bar10} and \cite{cub11} observed hot Jupiters by means of high-resolution spectroscopy at near-infrared wavelengths, but were not able to detect any molecules in their atmospheres.

Very recently, our group (Rodler et~al.~2012) as well as Brogi et al~(2012) reported the first successful detection of CO via high-resolution spectroscopic observations around 2.3~$\mu$m of the non-transiting hot Jupiter $\tau$~Boo~b. Both groups were able to measure the
orbital motion of this planet, which allowed them to determine the previously unknown orbital inclination of the planet and to finally solve for the exact planetary mass.

This paper is dedicated to the investigation of different data analysis approaches which were used by different research groups, with the goal to measure the planetary signal via high-resolution spectroscopy (Sections~2 and 3). In the second part of the paper, we present our studies of the search for carbon-monoxide in the atmosphere of the hot Jupiter HD~189733b (Section~4) and a brief summary of our results and conclusive remarks  (Section~5).

\section{Methods}
\subsection{Overview}
The idea of this approach is to observe spectral features in the planetary atmosphere via intermediate- and high-resolution spectroscopy. At visual wavelengths, the main contribution to the flux coming from a planet is the starlight reflected from the companion \citep{1998ApJ...502L.157S}. This means that the planet reflects the stellar spectrum, which is shifted in wavelength with respect to the star due to the orbital motion of the planet, and which is furthermore scaled down in intensity by a factor of several times $10^4$ due to the albedo of the planet, the illuminated fraction of the planetary disk visible, and the size of planet. Towards infrared wavelengths, the planet-to-star flux ratio drastically increases  to the order of $10^{-3}$ for hot Jupiters. At NIR wavelengths, the planetary flux is almost entirely produced by thermal emission from the planet. To measure and identify atmospheric features in the planetary atmospheres, theoretical models are required (e.g. \citealp{2007ApJS..168..140S}).

Key to this method is to observe a large number of spectral features in the planetary atmosphere to significantly overcome the planet-to-star flux ratio. Furthermore, it is important to take a time-series of spectra and therefore to observe the planetary features at different orbital phases of the planet, allowing to distinguish between the rather fixed stellar spectrum
 and the planetary one, which is periodically traveling in wavelength. 
 
 Data analysis involves the subtraction of the dominating stellar spectrum as well as of the telluric spectrum of the Earth's atmosphere at NIR wavelengths (see \citealp{1999ApJ...522L.145C,2002MNRAS.330..187C,Demi05,2007MNRAS.379.1097B}, and \citealp{2008A&A...485..859R}, for detailed descriptions). We note in passing that \citet{Lang11} reports a different approach by searching for the planet and stellar spectra in Fourier space.  
 
By adopting one of the methods described in Sections~\ref{s:deco} - \ref{s:data}, the spectral features of the planet in each residual spectrum  are then co-added to form a mean line profile of this spectrum. The following methods were used by different research groups: the deconvolution method, which was used by \citet{coli00,2002MNRAS.330..187C,2003MNRAS.344.1271L,2007MNRAS.379.1097B,barn07b,bar08,bar10}, and two straight-forward data modeling approaches adopting $\chi^2$-statistics (\citealp{1999ApJ...522L.145C,2008A&A...485..859R,rod10}) as well as  cross-correlation (\citealp{Sne10,Brog12,Rodl12b}). For each residual spectrum,  all three methods return -- in the ideal case -- a vector containing the mean line profile of the atmospheric features of the planet, which is Doppler shifted due to the instantaneous RV of the planet at the time of the observations (cf. Fig.~\ref{f5}).

The next step involves the alignment of all mean line profiles in the time series with the alignment being a function of the RV semi-amplitude of the planet $K_{\rm p}$. 
To this end, we transfer each of the planetary line profiles from the velocity grid relative to the star ($v$) to a grid based on $K_{\rm p}$:
\begin{equation} \label{e:1}
    K_{\rm{p}} = \frac{v}{\sin 2\pi\phi} ~,
  \end{equation} 
where $\phi$ is the orbital phase of the planet ($\phi=0$ occurs at mid-transit for transiting planets), which is a priori known from the RV solution of the star. 

The total of the aligned mean line profiles of the time series are then added up to form a single, overall mean line profile of the spectral features of the planet. We finally search for 
 the global minimum or the maximum peak, respectively, for $\chi^2$-statistics or cross-correlation / deconvolution method. Once such a candidate signal at a specific $K_{\rm p}$ has been found, the confidence level of the measurement is determined as described in Section~\ref{S24}.

One of the most important aspects of this technique is that it allows the determination of the orbital inclination $i$ of a non-transiting planet via $K_{\rm p}$ and
\begin{equation} \label{equ:doppler1}
 K_{\rm p}=K_{\rm{p,max}} \sin i 
  \end{equation} 
  and
\begin{equation} \label{2}
    K_{\rm{p,max}} = \Big(\frac{2\pi G~m_\star}{P_{\rm orb}}\Big)^{1/3} ~,
  \end{equation} 
 where $G$ is the gravitational constant, $m_\star$ the stellar mass, and $P_{\rm orb}$ the orbital period of the planet. 
 $K_{\rm{p,max}}$ is the maximum possible RV semi-amplitude that the planet can have, and which occurs for an orbital inclination $i=90^\circ$. The knowledge of the orbital inclination $i$ allows us to finally solve for the true planetary mass $m_{\rm p}=m_{\rm p,min} / \sin i$, where $m_{\rm p,min}$ denotes the minimum mass derived from the RV solution of the planet.  We note that Eq.~\ref{2} is only valid for $m_\star  \gg m_{\rm p}$.

\subsection{Deconvolution method}\label{s:deco}

The basic idea of this method is to deconvolve each observed star-free (and telluric-line-free) spectrum with a high-resolution reference spectrum that describes the atmospheric features in the planet, thereby summing up all these features into one mean line profile (e.g. \citealp{dona97,barn98}).

In general, the observed planetary spectrum $s(x)$ at a certain position $x$ on the
detector can be approximated as a convolution of the template spectrum $f(x')$  with the apparent mean
planetary line profile $p$. The template spectrum
$f(x')$ can be an artificial reference spectrum of very high resolution ($R>100,000$) or have the form of a list 
containing the positions and depths of the planetary absorption lines.

The observed planetary spectrum is given by 
\begin{equation} \label{E3:con}
   s(x) = \int\limits_{-\infty}^{\infty} f(x')~p(x-x')~{dx},
\end{equation}
with $\int\limits_{-\infty}^{\infty} p(x)~ dx = 1$. The discrete version of Eq.~\ref{E3:con} 
is
\begin{equation} \label{E3:con1}
   s_i = \sum\limits_{k=i-n}^{i+n} f_k~p_{i-k}\rm{,}
\end{equation}
where index $i$ is the pixel number of the observed object spectrum, and $k$ is the
index of the numerical grid used for the intrinsic spectrum \citep{Endl00}. $n$ denotes a cut-off parameter
of $p$ with $p_{i-k}=0$ for $|i-k|>n$, which allows us to shorten the infinite vector containing the mean line profile.

Following the algorithm by \citet{Endl00}, the grids of the reference spectrum $f$ and the mean profile $p$ can be oversampled
with respect to the grid on which the observed object spectrum is recorded.
The oversampled version of Eq.~\ref{E3:con1} follows as
\begin{equation} \label{E3:con3}
  s_i = \sum\limits_{j=qi}^{q(i+1)-1} (\sum\limits_{k'=j-nq}^{j+nq}
  f_{k'}~p_{j-k'}) \rm{,}
\end{equation}
where $q=k/i$ is the oversampling factor, and $j$ and $k'$ the indices of the
oversampled grids of the reference spectrum $f$ and the output vector $p$.  We combine the terms and rearrange Eq.~\ref{E3:con3} as follows:
\begin{equation} \label{E3:con4}
  s_i = \sum\limits_{k'=q}^{-q} (\sum\limits_{j=qi-k'}^{q(i+1)-1-k'} f_j)~ p_{k'}
\end{equation}
where $j$ is the index of the oversampled grid of the reference spectrum $f$, and $k'$
is the index of the
oversampled grid of the output vector $p$. Equation~\ref{E3:con4} is
nothing but a
matrix equation of the type
\begin{equation} \label{E3:con5}
  \overrightarrow{s} = \mathbf{F}~\overrightarrow{p}\rm{}
\end{equation}
with
\begin{equation} \label{E3:con46}
 \mathbf{F}_{ik'} = \sum\limits_{k=0}^{q-1}  f_{(q*i-k'+k)} \rm{.}
\end{equation}

Let $m$ be the dimension of the vector $\overrightarrow{s}$, which represents
the observed object spectrum, and $nq$ be the dimension of the oversampled vector
$\overrightarrow{p}$ containing the mean profile. Then the dimensions
of the matrix $\mathbf{F}$ are $m \times nq$. However, Eq.~\ref{E3:con5}
is incomplete since each data point $s_i$ of the observed data $\overrightarrow{s}$
exhibits its error $\Delta s_i$. The complete version of that equation follows as 
\begin{equation} \label{E3:con6}
  \overrightarrow{s} + \overrightarrow{\Delta s}= \mathbf{F}~\overrightarrow{p}\rm{.}
\end{equation}
Now we are ready for the calculation of the output vector $\overrightarrow{p}$
containing the planetary signal by using
a deconvolution algorithm; with this step, the signal-to-noise ratio (S/N)
of the planetary signal can be boosted by a factor $\sqrt{l}$ in the ideal case, where the $l$ spectral features have the same weight.

This problem constitutes an inversion problem which is ill-conditioned (due to $\overrightarrow{\Delta 
s}$), but over-determined (the size of the object spectrum is much larger than the size of the mean 
line profile). There are several algorithms to mathematically solve for this problem. We tested different deconvolution approaches and found that least-squares deconvolution preserves best the planetary signal. Thus, we modify Eq.~\ref{E3:con6} to form a least squares problem:
\begin{equation} \label{E3:lsd1}
\sum\limits_{i=1}^{m} \frac{(s_i -
  \mathbf{F}_{ik}~p_k)^2}{(\Delta s_i)^2}
 = (\overrightarrow{s} -
  \mathbf{F}~\overrightarrow{p})^T~\mathbf{E}~(\overrightarrow{s} -
  \mathbf{F}~\overrightarrow{p})
  \rightarrow \rm{min}\rm{,}
\end{equation}
where $\overrightarrow{\Delta s}$ is the error vector of $\overrightarrow{s}$, $\mathbf{E}=\rm{Diag}[\Delta s_1^{-2}, ...,\Delta s_m^{-2}]$ and $m$ is
  the dimension of the vector of the observed object spectrum
  $\overrightarrow{s}$. 
  We find the least squares solution for the output vector containing the mean line profile of the planet
  $\overrightarrow{p}$ by solving the matrix equation obtained by multiplying
  both sides of Equation~\ref{E3:con5} with ~$\mathbf{F}^T~\mathbf{E}$:
\begin{equation} \label{E3:lsd2}
\mathbf{F}^T~\mathbf{E}~\overrightarrow{s}
=\mathbf{F}^T~\mathbf{E}~\mathbf{F}~\overrightarrow{p} 
\end{equation}
\begin{equation} \label{E3:lsd3}
\Rightarrow
\overrightarrow{p}=(\mathbf{F}^T~\mathbf{E}~\mathbf{F})^{-1}~\mathbf{F}^T~\mathbf{E}~\overrightarrow{s} \rm{.}
\end{equation}
The matrix resulting from the multiplication
$\mathbf{F}^T~\mathbf{E}~\mathbf{F}$ is square, symmetric and positive
definite. Therefore, the inverse matrix can be calculated by Cholesky
decomposition (Press et al.~1992).

\subsection{Data modeling and $\chi^2$-statistics}\label{s:chi2}

For each of the residual spectra (i.e. stellar-free and telluric-free spectra), we create a planetary model $M$ being a theoretical spectrum for the planetary atmosphere $f$ (e.g. \citealp{2007ApJS..168..140S}), which is degraded in spectral resolution to the resolution of the observations. This version of the planetary model $M$ is then Doppler-shifted as a function of $K_{\rm p} \sin 2\pi\phi$ (remember that $K_{\rm p}$ is the RV semi-amplitude of the planet, which is unknown for most of the non-transiting planets; $\phi$ is the orbital phase of the planet which is a priori known from the RV solution of the star) with respect to the star and interpolated on the pixel grid of the observations. The spectrum is furthermore scaled in intensity as a function of a planet-to-star flux ratio ($\epsilon_{\lambda}$) at wavelength $\lambda$ as well as to the phase-function (i.e. the illumination geometry of the observed planetary disk at different orbital phases  - see e.g. \citealp{2008A&A...485..859R}). For pixel $k$, 
 \begin{equation} \label{E4.2a}
    M_{k} = \epsilon_{\rm p} ~f_{k} \big\lbrace \lambda_k~(1+K_{\rm p} \sin 2\pi\phi~~c^{-1})\big\rbrace {\rm,}
 \end{equation}
  where $c$ denotes the speed of light.

Varying the free parameters $K_{\rm p}$ and $\epsilon$,
  we finally search for the best-fit model $M$ to all the residual spectra $s$
   by way
  of $\chi^2$-minimization in appropriate search ranges, where the reduced $\chi^2$ is
   \begin{equation} \label{E4.2b}
\chi^2_\nu =\frac{1}{N-m}\sum\limits_{k} \frac{(M_k-s_k)^2}{\Delta s_k^2}\rightarrow \rm{min}\rm{.}
 \end{equation}
$N$ is the total number of data points, and $m$ is the number of fitted parameters. Note that $\Delta s$ denotes the errors of $s$.

\subsection{Data modeling and cross-correlation}\label{s:data}
 
 For each residual spectrum $s$, we Doppler-shift the atmospheric model spectrum $f$ of the planet by velocity $v$ (see Eq.~1) in a given search range and interpolate it onto the pixel grid of $s$. We then calculate the normalized correlation degree $C(v)$ for each residual spectrum following the formalism by \citet{cub11}:
 \begin{equation} \label{E4.2c}
    C(v) = \frac{w}{W} \frac{\sum\limits_{k}\Big\lbrace\big(f_{k}(v)-\bar{f}(v)\big)\big(s_{k}-\bar{s}\big)\Big\rbrace}{\sqrt{\Big\lbrace\sum\limits_{k}\big(f_{k}(v)-\bar{f}(v)\big)^2 \Big\rbrace\Big\lbrace\sum\limits_{k}\big(s_{k}-\bar{s}\big)^2\Big\rbrace}}~,
     \end{equation}
where index $k$ denotes the pixel number, $\bar{f}$ and $\bar{s}$ are the mean values of the atmospheric model spectrum and residual spectrum, respectively. The parameter $w$ denotes the weight for the specific residual spectrum, while $W$ is the sum of the weights of all residual spectra which are used in the cross-correlation.

  \subsection{Determination of the confidence level}
\label{S24}

\begin{figure}
   \includegraphics[angle=0,width=8.6cm]{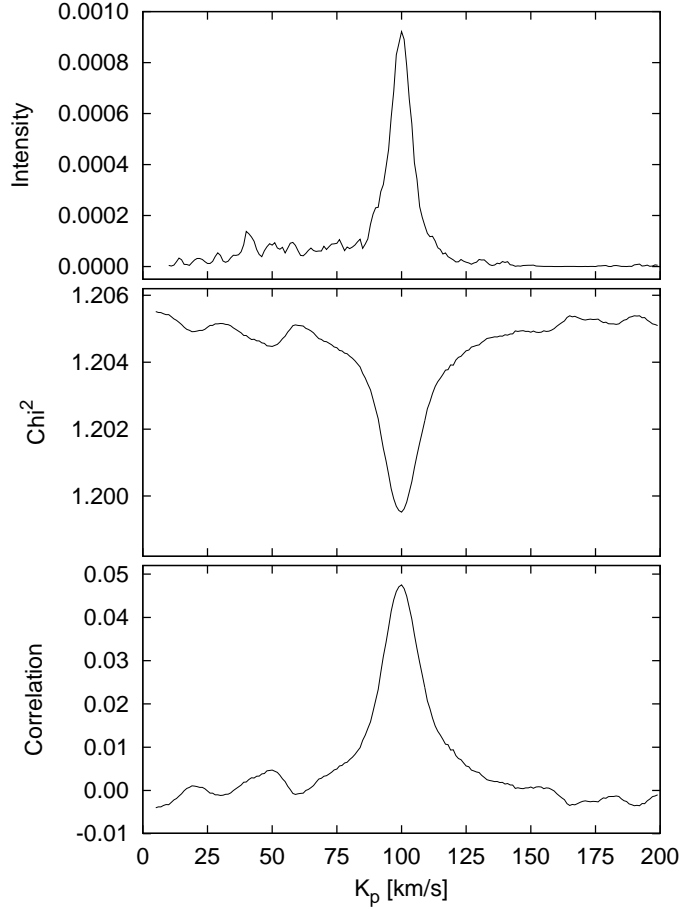}
\caption{Output of the different data analysis methods for a single spectrum:
 deconvolution method (uppermost panel), data modeling with $\chi^2$-statistics (middle) and cross-correlation (bottom). The data to be analyzed had been created in the way described in Section~\ref{S12} by adopting a scaling value of $\epsilon=10^{-3}$ and an RV semi-amplitude of $K_{\rm p}=100$~km~s$^{-1}$.
 }
\label{f5}
\end{figure}

Once a candidate signal of the planetary spectrum has been located, its confidence level needs to be determined. Note that this candidate feature is a peak in the case where the deconvolution method or the cross-correlation analysis has been employed. If the data was analyzed adopting $\chi^2$-statistics, the candidate signal is reverted and produces a dip (Fig.~\ref{f5}). 

One way to determine the confidence level of the candidate signal is to employ a
   bootstrap randomization method (e.g. \citealp{1997A&A...320..831K}): 
   random values of the orbital phases are assigned to the observed spectra, thereby creating $N$ different data sets ($N=100,000$ in our analyses).
   Any signal present in the original data is then scrambled in these
   artificial data sets.
   For all these randomized data sets, we re-run the data analysis in given parameter search ranges and locate the candidate feature with its maximum value or its minimum $\chi^2_{\nu,\rm min}$ value, depending on the employed method.
   
The confidence level of the candidate features is estimated to be $\approx 1-{\rm FAP} = 1-m/N$, where FAP is the false alarm probability, $m$ is the number of the best fit models  having a $\chi^2_\nu$ value smaller or equal than the $\chi^2_{\nu,\rm min}$ value found in the original, unscrambled
   data sets. We note that for the other two data analysis methods, we count the number of the best fit models yielding an equal or higher peak than the peak value of the candidate signal. 

  \subsection{Determination of the error range of the result}
\label{S26}
In the case of a significant detection of the planetary atmosphere ($\ge 3\sigma$, which corresponds to $\ge 0.9973$ confidence), we determine the $1\sigma$-error of the free parameters $K_{\rm p}$ and $\epsilon$ via bootstrap resampling \citep{bar92}.
To this end, we create randomly resampled data sets of the size of the original data set. 
In the resampling process, each set of spectra is randomly drawn from the general
pool of spectra. For each spectrum a copy is kept for the artificial data set, but the
original spectrum is returned to the pool with the effect that it can be drawn again.
This way some of the original spectra will not appear in an artificial data set while
others may appear more than once.
Then, the parameters $K_{\rm p}$ and $\epsilon$ of the best fit model are stored. We create a total of 10,000 artificial data sets to get a more precise estimate of the distribution of $K_{\rm p}$ (and $\epsilon$).

 To determine the 1$\sigma$-errors of the model parameters $K_{\rm p}$ and $\epsilon$ for the
original data set we examine the distributions that were obtained for these
parameters for the artificial data sets.  We create 68.3\% confidence intervals around
the original  $K_{\rm p}$ and $\epsilon$ values by taking the ranges adjacent to either side of the
original values that each contain 34.15\% of the artificial values.

\section{Simulations}
\subsection{Data}\label{S12}
\begin{figure}
   \includegraphics[angle=-90,width=8.6cm]{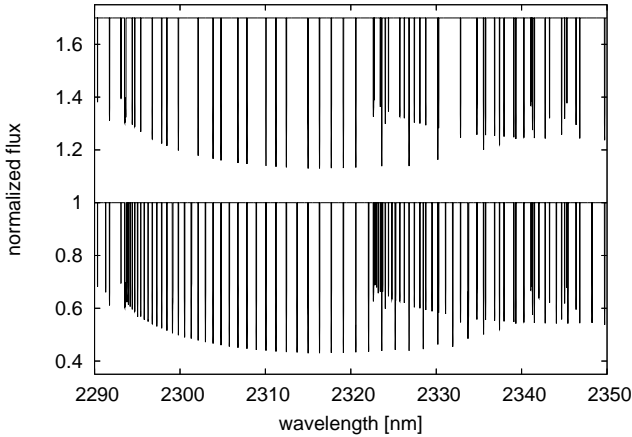}
\caption{Reference spectra adopted for the data analysis. The lower spectrum depicts a theoretical PHOENIX model for a brown dwarf having a temperature of $T=1800$~K and solar metallicity. We used this spectrum to create the data sets as well as for data analysis. In the upper spectrum, the same spectrum is shown, but we removed 20\% of the lines to study also the effect of missing lines in the data analysis (cf. Case~2 in the text). We note that this normalized spectrum has been shifted up for clarity.}
\label{f4}
\end{figure}

We created artificial data sets with the goal of testing the three data analysis methods and to find out which is most sensitive for the task. We adopted a PHOENIX model spectrum  \citep{1997ApJ...483..390H,2008ApJ...675L.105H,2011A&A...529A..44W} for a brown dwarf having a temperature of $T=1800$~K and solar metallicity around 2.3~$\mu$m (Fig.~\ref{f4}, lower spectrum). 
In this wavelength regime, the brown dwarf spectrum exhibits a dense forest of absorption lines
almost entirely produced by the molecule carbon monoxide (CO).

Each data set consisted of 79 spectra, which were Doppler-shifted as follows.
We assigned an orbital phase value to each of the spectra, starting with $\phi=0.305$ for spectrum 1 and subsequently increasing the value of $\phi$ by 0.005 for the following spectra. In total, the orbital phases  ranged from $\phi=0.305$ to $0.695$ and therefore comprised the regions where a planet would appear at its brightest.
Since hot Jupiters typically have RV semi-amplitudes of the order of $K_{\rm p}=100$~km~s$^{-1}$, we chose $K_{\rm p}=100$~km~s$^{-1}$ and calculated the value of the instantaneous RV shift $v$ of the model spectrum by Eq.~\ref{e:1}. 

We simulated the data for a cross-dispersed infrared spectrograph of high resolution. Thus, we degraded the spectral resolution of the spectrum to $R=60,000$ by a convolution of the model spectrum with a suitable Gaussian and interpolated the resulting spectrum onto a pixel grid ranging from 2.3 to 2.35~$\mu$m. This pixel grid consisted of a total of 4800~px, and one pixel corresponded to a velocity range of 1.5~km~s$^{-1}$. We then scaled the depths of absorption lines in the spectra due a chosen star-to-planet flux ratio $\epsilon$ between $10^{-3}$ to $5\times10^{-5}$. 
In the final step, we added Poisson noise to the data in such a way that the S/N level was at 500 per spectral pixel, which is a typical value for high-resolution spectra in the infrared.

We analyzed these artificial data sets by the three data analysis methods and investigated also the three following cases: In Case~1, we analyzed the data by adopting the PHOENIX model spectrum which we had used for creating the data sets. For the analysis with the data modeling approaches, we further degraded the spectral resolution of the reference spectrum to 60,000.

In Case~2, we explored the influence of missing lines in the reference spectrum, which was then adopted for the data analysis. The reference spectrum was a version of the brown dwarf spectrum, in which we had randomly removed 20\% of the lines (Fig.~\ref{f4}, upper spectrum).

In the last case (Case~3), we studied how tiny continuum variations affected the result of the data analysis. To this end, we added a continuous sine wave to the modeled spectra having a period of 1000 pixel and a semi-amplitude of $10^{-3}$. Data analysis was carried out by adopting the PHOENIX model spectrum containing all absorption lines.

\subsection{Results}\label{S12a}

\begin{figure}
   \includegraphics[angle=-90,width=8.6cm]{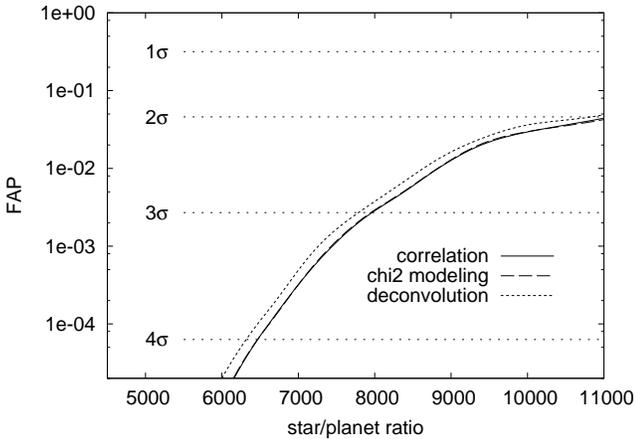}
\caption{Comparison between the data analysis methods for Case~1. We created different data sets for different planet-to-star flux ratios $\epsilon$ (for clarity, we show the inverse of $\epsilon$) and analyzed them with the three data analysis methods. The solid, dashed and dotted graphs depict the false alarm probability (FAP) of the results obtained with the deconvolution method, the data modeling method using $\chi^2$-statistics, and with cross-correlation, respectively.
 The four dotted horizontal lines mark the confidence levels in $\sigma$-units. As we compare these results, we find that all methods are almost equally sensitive, with cross-correlation and $\chi^2$-data modeling showing systematically a higher sensitivity than the deconvolution method.}
\label{f3}
\end{figure}

\begin{figure}
   \includegraphics[angle=-90,width=8.6cm]{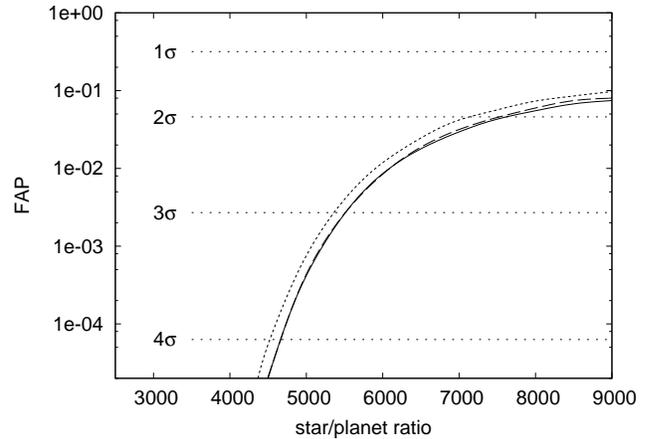}
\caption{Same as in Fig.~\ref{f3}, but for Case~2. We investigated how an incomplete spectral
information in the reference spectrum affects the data analysis. The data set was created with the correct brown dwarf model spectrum, while for the data analysis we adopted a version of that model spectrum, in which we had deleted 20\% of the lines (cf. Fig.~\ref{f4}). In comparison to the other cases (Figures~\ref{f3} and \ref{f1}), we see that the sensitivity of all three methods is strongly affected by an incomplete reference spectrum.}
\label{f2}
\end{figure}

\begin{figure}
   \includegraphics[angle=-90,width=8.6cm]{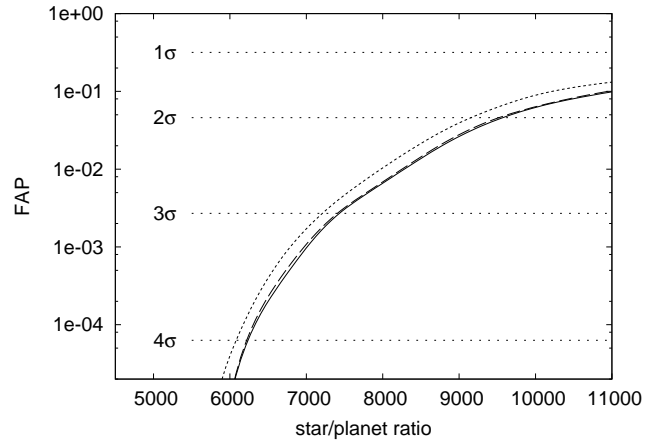}
\caption{Same as in Fig.~\ref{f3}, but for Case~3. We investigated how a variable continuum affects  the data analysis. To this end, we added a sine wave to the data having a period of 1000 pixel and a semi-amplitude of $10^{-3}$. As result, we again find that the deconvolution method is significantly less sensitive than the other two data analysis methods.}
\label{f1}
\end{figure}

For all three cases, we find that all three methods are almost equally sensitive, with cross-correlation and $\chi^2$-data modeling showing systematically a higher sensitivity than the deconvolution method (Figures~\ref{f3} - \ref{f1}).

For Case~1 (CO spectra), we are able to retrieve the planetary signal at the correct value of $K_{\rm p}$ at the $3\sigma$-confidence level down to planet-to-star flux ratios of $\epsilon\approx1/8000$ for all three methods. 

For Case~2 (fewer lines), we made use of the same data sets which had been created for Case~1, and analyzed them with the model spectrum for which we had randomly had removed 20\% of the absorption lines.  Fig.~\ref{f2} illustrates that by employing cross-correlation, the data modeling approach using $\chi^2$-statistics, and the deconvolution method, the planetary signal with a planet-to-star flux ratio of $\epsilon\approx1/5500$ is retrieved at the $3\sigma$-confidence level. 

In Case~3 (variable continuum),  we adopted the same data sets which we had created for Case~1, but multiplied the spectra with a sine wave to simulate variable continuum. Data analyses were carried out by adopting the correct spectrum of the brown dwarf. As shown in Fig.~\ref{f1}, we are able to retrieve the planetary signal at the correct position and at the $3\sigma$-confidence level down to planet-to-star flux ratios of $\epsilon\approx1/7500$ with all three data analysis methods. 

As we compare the individual methods and cases, we realize that the deconvolution method is systematically the least sensitive one of the three data analysis techniques. We attribute the $\approx 3-7\%$ lower sensitivity of the deconvolution approach to the additional processing layer, where a mean line profile is calculated at once according to a mathematical constraint (in our case: least-squares minimization). As a second point, for the deconvolution we adopt a reference spectrum consisting of delta functions (at the reference line positions), which is sampled onto the same pixel grid as the observed spectrum. Due to that discrete sampling of the reference spectrum, the position of each reference line is at a full pixel, i.e. the center of the line is likely to be shifted by a fraction of a pixel. This, and the common case of line blending, where two lines might be treated as one, might produce distorted line profiles which then affect the overall mean line profile.

\section{Reanalysis of the HD~189733b-NIRSPEC data set}
\subsection{The planet HD~189733b}

The parameters of HD~189733b and its parent star are provided in Table~\ref{tab:tauboo}.
This transiting planet is among the best studied exoplanets so far. Discovered in 2005 by \citet{Bouc05}, it has quickly become the favorite target for planet atmosphere studies, being located in one of the brightest known transiting system. Key results include the detection of a hot spot on the planet surface \citep{Knut07,Knut12} by studying the phase function of the planet in the infrared, indicating a temperature around 1300~K, as well as the discovery of high-altitude haze (\citealp{Pont08,Sing11}). Several atoms and molecules in the planet atmosphere were found: water (\citealp{2008Natur.456..767G}), sodium seen in absorption at visual wavelengths \citep{Redf08}, as well as methane and carbon dioxide (e.g. \citealp{2009ApJ...690L.114S,2009ApJ...699..478D,Wald12}). \citet{2009ApJ...699..478D} found strong absorption around 4.5~$\mu$m, probably due to CO.  \citet{Leca10} measured strong evaporation of the planetary atmosphere due to the high irradiation from the host star. In addition to these discoveries, the spectrum of the dayside emission has been measured at NIR- and mid-infrared wavelengths (\citealp{Char08,2008Natur.456..767G}; and \citealp{Wald12}). 

 \begin{table}
    \caption{Parameters of the star HD~189733 and its planetary
  companion. Abbreviations for the references are: 
     Bak6~=~\citet{Bako06}.
     Bar7~=~\citet{2007MNRAS.379.1097B}.
     Bou5~=~\citet{Bouc05}.
     Knu7~=~\citet{Knut07}.
     VV9~=~\citet{2009ApJ...694.1085V}.}             
    \label{tab:tauboo}      
    \begin{center}                         
    \begin{tabular}{l r l l }        
      \hline\hline                 
      Parameter & Value & Error &Ref. \\
      \hline                        
      Star: \\
      Spectral type & K1-2 V  & & Bou5 \\ 
      $K~(mag)$        & 5.54 & 0.02 & VV9\\
      $m_{\star}~ (\rm{M_{\odot}})$ & 0.82 & 0.03 & Bou5 \\
      $R_{\star}~(\rm{R_{\odot}})$ & 0.76 & 0.01 & Bou5 \\
      $T_{\rm eff}$~(K) & 5000 & 50 & Bou5 \\
      \hline
      Planet:\\
      $m_{\rm{p}} ~(\rm{M_{\rm{Jup}}})$  & 1.15 & 0.04 & Bou5 \\
      $R_{\rm{p}} ~(\rm{R_{\rm{Jup}}})$  & 1.154 & 0.033 & Bak6 \\
      ${T_{\rm eff}}$ (K) & 1300  & 200  & Knu7 \\
      $a~ (\rm{AU})$ & 0.0313 & 0.0004 & Bou5 \\
      $i~ (^\circ)$ & 85.79  & 0.24 & Bak6 \\      
      $P_{\rm orb}$ (d) & 2.218~5733 & 0.000~0019  & Bak6 \\
      ${t_{\phi=0}}$ (BJD) & 245~3988.803362  & 0.0023  & Bak6 \\
      $K_{\rm p}~(\rm{km~s^{-1}})$ & 152.6 & 2.0 & Bar7 \\
      \hline                                   
    \end{tabular}
    \end{center}
  \end{table}

\subsection{NIRSPEC data and their reanalysis}
We reanalyzed the data set published in \cite{bar10}, which had been taken with the goal to detect H$_2$O and CO in the atmosphere of HD~189733b. Their data analysis, however, resulted in a detection  of low significance (98.8\%) of these elements in the planetary atmosphere. Data were obtained with NIRSPEC \citep{Mcle98} at the Keck II Telescope, Hawaii, USA, on UT 2008 June 15 and June 22, when the dayside of the planet was almost entirely visible. A total of 373 spectra were recorded using a $1024\times1024$ InSb Aladdin-3 array. The spectra were taken with a slit width of 0.432 arcsec, giving a spectral resolution of $R\approx 25,000$. 

\begin{figure}
   \includegraphics[angle=-90,width=8.6cm]{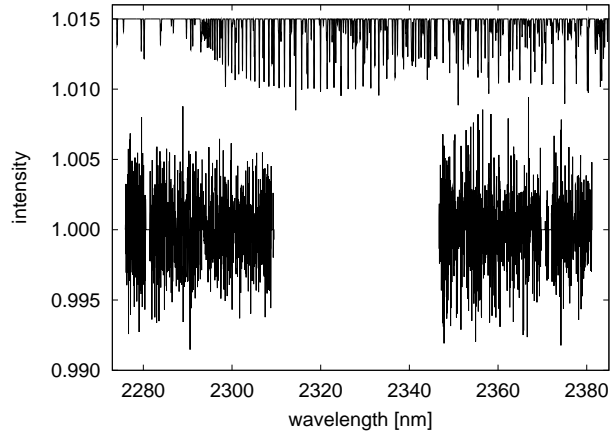}
\caption{Upper spectrum: the CO model with spectral resolution of 25,000. For clarity, the normalized spectrum is shifted up by 0.005 and scaled for a planet-to-star flux ratio of $\epsilon=10^{-2}$, which is  a factor 4.5 larger than the actually measured value by \citet{Wald12}. Lower spectrum: the two spectral orders used in the data analysis. The  residual spectra are shown after the subtraction of the stellar- and telluric absorption lines as well as after the bad pixel removal. }
\label{f7}
\end{figure}

Using the method outlined in \citet{2007MNRAS.379.1097B} and \citet{bar08}, we reduced the data and attempted to extract the planetary signature from time-series spectra by removing the dominant spectral contributions: namely the stellar spectrum and the telluric lines. Contrary to the data analysis of Barnes et al.~(2010), we restricted the data analysis to the last two spectral orders comprising the wavelength region of $\lambda = 2.275$ to $2.31~\mu$m and $\lambda = 2.347$ to $2.383~\mu$m, respectively (Fig.~\ref{f7}), where we expected the dense CO absorption forest of the companion spectrum.  We note that the latter spectral order was not used in the data analysis by \cite{bar10}.
In the wavelength regime of those two spectral orders, \citet{Wald12} reported a planet-to-star flux ratio of $\epsilon= 2.2\times10^{-3}\approx 1/450$ from secondary eclipse measurements of HD~189733b.

\begin{figure}
   \includegraphics[angle=-90,width=8.6cm]{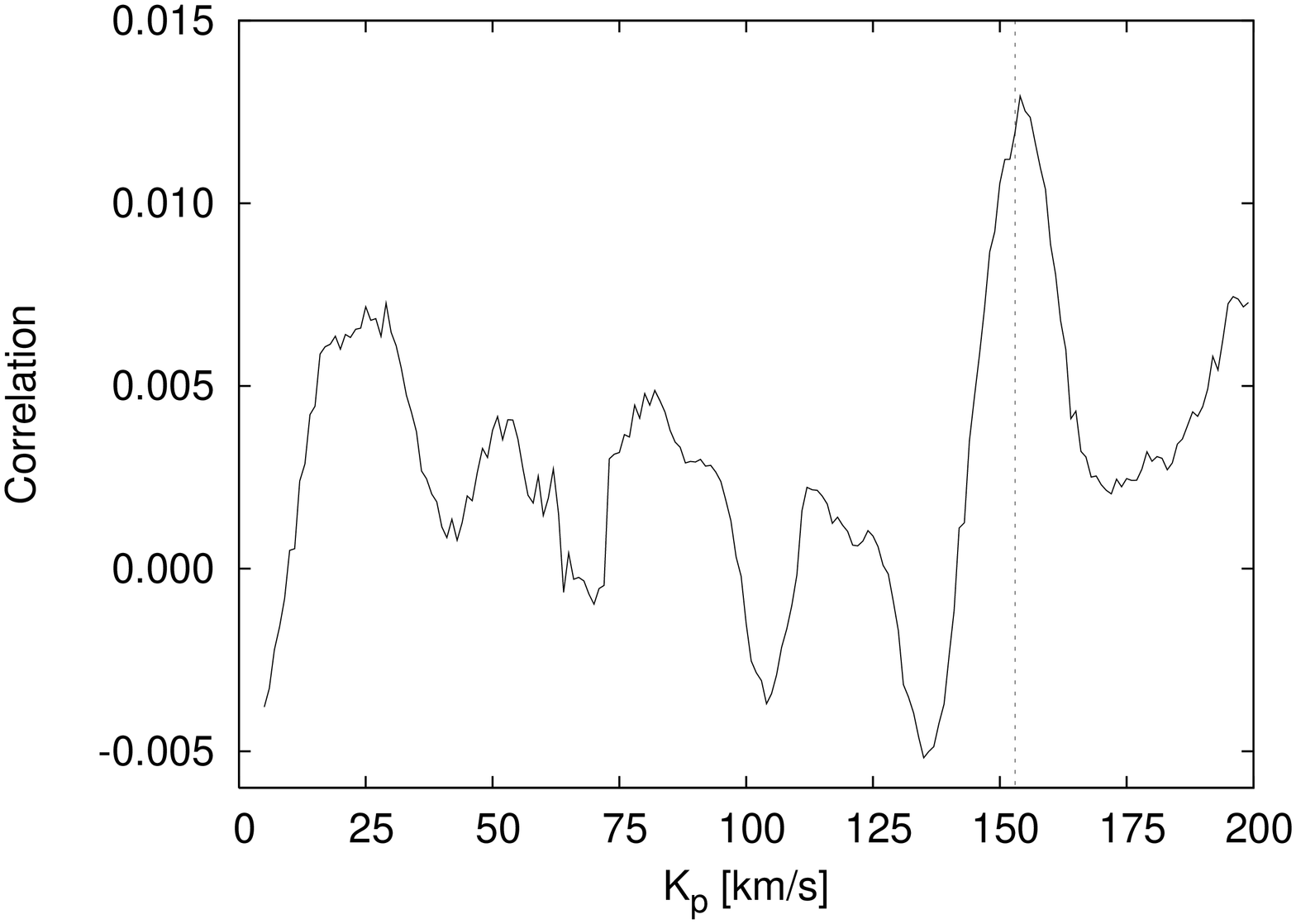}
   \includegraphics[angle=-90,width=8.6cm]{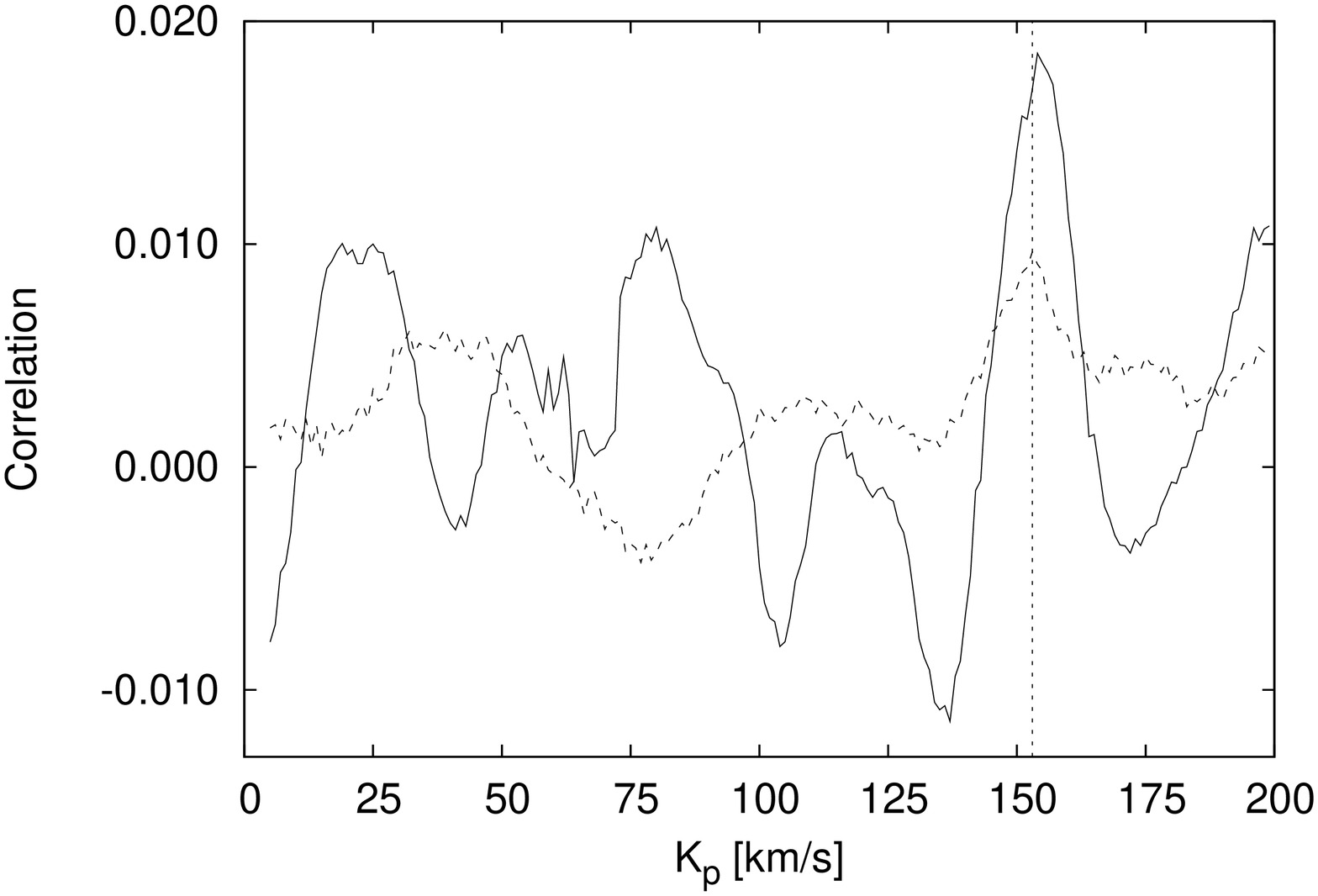}
\caption{Upper panel: co-added cross-correlation functions in the rest-frame of the planet. The peak of the CCF occurs at an RV semi-amplitude of $154$~km~s$^{-1}$ and appears close to the known RV semi-amplitude of HD~189733b being $K_{\rm p}=152.6$~km~s$^{-1}$ (dotted vertical line). While bootstrap randomization runs showed that the peak of the CCF is at the confidence level of 99.54\% ($2.9\sigma$), a more straight-forward approach revealed that the candidate feature is significant with 99.92\% ($3.4\sigma$) confidence and therefore represents a detection of the CO absorption line forest in the planetary atmosphere spectrum of HD~189733b. Lower panel: same as above, but showing the results of the cross-correlation functions for the first night (solid line) and second night (dotted line).}
\label{f6}
\end{figure}

\begin{figure}
   \includegraphics[angle=-90,width=8.6cm]{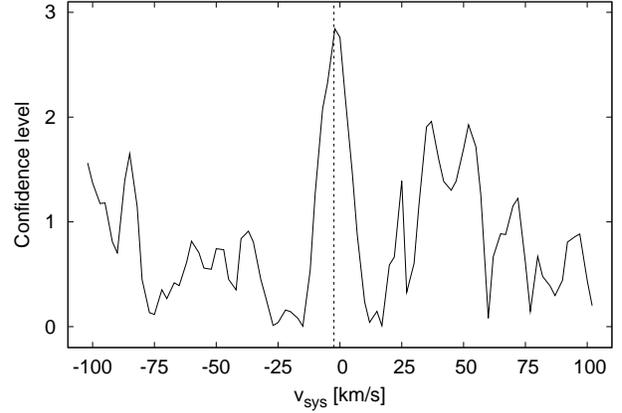}
\caption{As a test, we varied the systemic RV of the star HD~189733 and determined the confidence levels at the measured RV semi-amplitude of the planet, $K_{\rm p}=154$~km~s$^{-1}$. The confidence levels had been determined by bootstrap randomization runs.
A clear signal at the systemic velocity of HD~189733. being $v_{\rm sys}$=-2.4~ km s$^{-1}$ (dotted line) is visible.}
\label{f15}
\end{figure}

\begin{figure}
   \includegraphics[angle=-90,width=8.6cm]{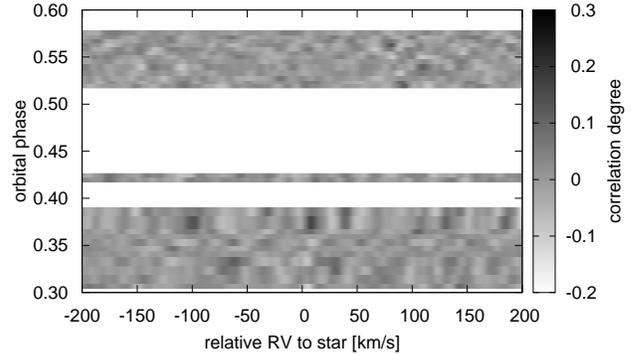}
\caption{The individual cross-correlation functions (CCFs) of the 269 residual spectra of HD~189733b with the CO model spectrum in the rest-frame of the star are shown. During the course of the observations, the orbital motion of the planet produces a RV-shift starting at about 150~km~s$^{-1}$ and ending at -75~km~s$^{-1}$, respectively for  orbital phases of 0.30  and 0.58. However, no trace of the planet appears. The linear grey-scales indicate the strength of the cross-correlation signal (dark means absorption, bright means emission).}
\label{f12}
\end{figure}

\begin{figure}
   \includegraphics[angle=-90,width=8.6cm]{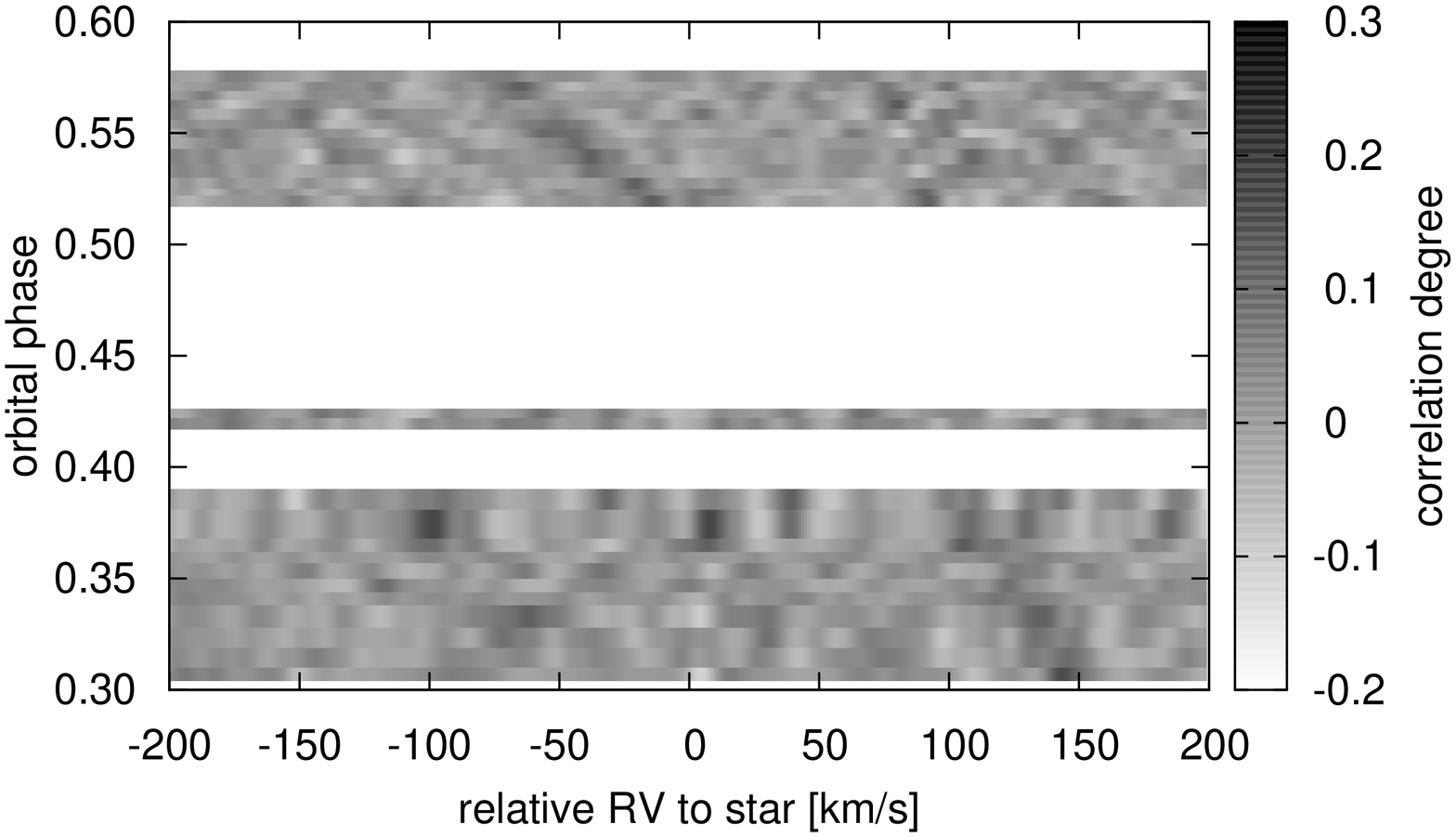}
\caption{We injected a spectrum showing CO with solar abundance, with a spectral resolution of $R=25,000$ and for planet-to-star flux ratio of $1/450$ to the residual spectra of  HD~189733b. The individual cross-correlation functions (CCFs) of the 269 spectra, to which  we injected a planetary signal, with the CO model spectrum having solar abundance in the rest-frame of the star are shown. The linear grey-scales indicate the strength of the cross-correlation signal (dark means absorption, bright means emission). It is shown that  the trace of the RVs of the injected planetary signal can be recovered. }
\label{f10}
\end{figure}


The S/N of the residual spectra (i.e. after the removal of the stellar spectrum and telluric lines) was on average 370 and 450 per spectral pixel in the first and second night, respectively. We rejected those frames taken when the planet was behind the star and not visible. In the end, we worked on 153 and 116 residual spectra, almost entirely covering the orbital phases $\phi=0.302$ to $0.421$ and $\phi=0.515$ to $0.580$, respectively for the first and second night, when the day-side of the planet was largely visible. We further identified bad pixels and outliers and discarded them from the data analysis. We adopted cross-correlation (Section~2.4) to search for the CO spectrum of the planet in the residual spectra. As reference spectrum, we adopted a PHOENIX spectrum of a brown dwarf with a temperature of T=1500~K and with a spectral resolution of $R=25,000$ (Fig.~\ref{f7}).  In the cross-correlation, we weighed the spectra according to their average SNR level per spectral pixel, and further accounted for both the systemic radial velocity of the star HD~189733 ($v=-2.4$~km~s$^{-1}$) and the barycentric velocity of the Earth. We then co-aligned and co-added the individual cross-correlation functions in the planet's rest-frame, thereby taking into account the orbital phase information of the planet at the barycentric Julian date of the observations.

\subsection{Results and discussion}
As shown in Figure~\ref{f6}, we find the strongest candidate feature at $K_{\rm p}=154$~km~s$^{-1}$, which is located within the 1$\sigma$-error range of the known RV semi-amplitude of HD~189733b being, $K_{\rm p}=152.6\pm 2$~km~s$^{-1}$. 
 
Since the RV semi-amplitude of the planet $K_{\rm p}$ had already  been estimated by \citet{barn07b}, we can restrict the bootstrap randomization run to the search range $K_{\rm p}=152.6\pm3\sigma$~km~s$^{-1}$ (i.e. $146.6$ to $158.6$~km~s$^{-1}$). As result of this bootstrap randomization run, we found that the candidate feature is at a confidence level of $99.54\%$ and therefore represents a $2.9\sigma$-detection of CO in the planetary atmosphere of HD~189733b. 

We furthermore carried out the data analysis on the two independent nights and found this candidate feature present in both nights (Fig.~\ref{f6}, lower panel). As a plausibility test, we varied the systemic RV of HD~189733 and determined the confidence levels at the measured RV semi-amplitude of the planet, $K_{\rm p}=154$~km~s$^{-1}$. The highest peak occurs at the genuine system velocity of HD~189733, which is $v_{\rm sys}$=-2.4~ km s$^{-1}$ (Fig.~\ref{f15}). 

In addition to that, we adopted a different strategy to determine the confidence level of the candidate feature.
To this end, we calculated the cross-correlation values for all combinations of the two parameters $K_{\rm p}$ and the systemic velocity $v_{\rm sys}$ of HD~189733b, respectively, in the range of $5 \le K_{\rm p} \le 200$~km~s$^{-1}$ and $-100 \le v_{\rm sys} \le 100$ ~km~s$^{-1}$. The average value of the cross-correlation values for all those combinations is 0.00055, its scatter (rms) is 0.00368, while the peak value of the candidate feature shows 0.0129. We subtracted the average value from the peak value and divided the result by the scatter of the cross-correlation functions. This revealed that the candidate feature is indeed significant at the 99.92\% ($3.36\sigma$)-confidence level.

When plotting the individual cross-correlation functions  of the residual spectra with the CO model spectra, we do not see the planetary signature (Fig.~\ref{f12}). Given the relatively large flux ratio between the dayside of HD~189733b and its host star, we should have been able to measure the trace of the planetary RV signal. Albeit the presence of systematic noise that hamper the retrieval of the weak planetary signature, that may suggest low abundance of CO in the planetary atmosphere of HD~189733b.

To demonstrate the power of this approach, we injected an artificial planetary signal into the residual spectra and analyzed these data sets with cross-correlation. We modeled the artificial planetary signal adopting CO spectra with solar abundance with a spectral resolution of  $R=25,000$. The spectra were then scaled to an intensity ratio of 1/450 and shifted in RV according to the orbital motion of HD~189733b (i.e. for an RV semi-amplitude of $K_{\rm p}= 153$~km~s$^{-1}$). The results of the data analysis are shown in Figure~\ref{f10}. The injected planetary signal could be recovered at the correct value of $K_{\rm p}$, and bootstrap randomization simulations for a search range of $K_{\rm p}$ from 5 to 200~km~s$^{-1}$ revealed that the signal is significant at  a confidence level of $>99.99$. By adopting bootstrap resampling simulations, we furthermore determined the $1\sigma$-error ranges of the recovered planetary signal, being 2.5~km~s$^{-1}$. Figs.~\ref{f12} and \ref{f10} also show some strong cross-correlation artifacts which are of the order of the injected planetary signal. We attribute these artifacts mainly to systematic errors coming from the removal of the stellar signal and the telluric spectrum. 

To estimate the flux ratio between the strong planetary lines $F_{\rm lines}$ and the stellar continuum $F_\star$, we again injected an artificial planetary signal into a scrambled set of the residual spectra and analyzed these data sets with cross-correlation. The injected planetary spectrum was scaled to a chosen intensity ratio $F_{\rm lines}/F_\star$ and Doppler-shifted as described above. We recovered the injected the planetary signal and determined its confidence level by the aforementioned straight-forward strategy. We found that for a scaling factor of $F_{\rm lines}/F_\star=1.8\times10^{-4}$, we attain $3.4\sigma$ confidence. This value again indicates a low abundance of CO in the planetary atmosphere of HD~189733b, given the large flux ratio between the continua of the planet day-side spectrum
and the stellar one, being $\epsilon=2.2\times10^{-3}$ around 2.3~$\mu$m.



\section{Summary and Conclusions}
We carried out studies to find out what data analysis approach is best suited for the search for atoms and molecules in hot Jupiters via high-resolution spectroscopy. We first created artificial data sets consisting of spectra of planetary atmospheres, scaled them in intensity according to a chosen value of $F_{\rm lines} / F_{\star}$, and analyzed them by different data analyses approaches. As result, we found that the highest sensitivities to recover the weak planetary features are attained with cross-correlation and $\chi^2$-data modeling, while the deconvolution method was less sensitive ($\approx3-7\%$) than the two aforementioned methods.

In light of these studies, we attempted to measure the dense CO absorption line forest around 2.3~micron in the day-side spectrum of the transiting hot Jupiter HD~189733b. By employing cross-correlation, we reanalyzed a time-series of spectra taken with the Near Infrared Spectrometer (NIRSPEC) at Keck II during two nights and detect a candidate planet signal at an RV semi-amplitude $K_{\rm p}=154$~km~s$^{-1}$, which is located within the 1$\sigma$-error range of the known RV semi-amplitude of HD~189733b ($K_{\rm p}=152.6\pm2$~km~s$^{-1}$). While bootstrap randomization runs resulted in $2.9\sigma$-confidence for this candidate feature, a more straight-forward 
test revealed that we detect the planetary signal with a S/N of $3.4$.

As a plausibility test, we independently carried out the data analysis for each of the two nights and found this candidate feature clearly present in both nights. In addition, we varied the systemic RV of HD~189733 and found the highest peak at the genuine system velocity of HD~189733, which is $v_{\rm sys}$=-2.4~ km s$^{-1}$. 

In the past, \citet{bar10} analyzed the same data set in the wavelength region 2.21 - 2.36~$\mu$m and searched for water and CO absorption in the planetary atmosphere. These authors adopted a different data analysis strategy (different bad pixel correction, deconvolution method, different spectral orders used) for their purposes and found in their data analysis a candidate feature of the planetary signal close to the RV semi-amplitude of the planet at the  98.8\% confidence level. 

We are consequently confident to claim a detection of CO absorption in the planetary atmosphere of HD~189733b. This work demonstrates the power of intermediate-resolution spectroscopy at infrared wavelengths to investigate the atmospheres of remote planets. The measured planetary CO signal is weak and may suggest a low abundance of CO in the planetary atmosphere of HD~189733b. In addition, we observe CO in absorption, which indicates that the atmosphere of HD~189733b lacks a strong thermal inversion layer.
 

\section*{Acknowledgments}
We thank those of the Hawaiian ancestry on whose sacred mountain we are privileged to be guests. JB gratefully acknowledges funding through a University of Hertfordshire Research Fellowship. During this research, FR and JB have received travel support from RoPACS, a Marie Curie Initial Training Network funded by the European Commission's Seventh Framework Programme. We furthermore would like to thank the anonymous referee for very helpful and constructive comments.

\bsp

\label{lastpage}


\begin{thebibliography}{39}
\expandafter\ifx\csname natexlab\endcsname\relax\defNatureexlab#1{#1}\fi
\expandafter\ifx\csname href\endcsname\relax
  \def\href#1#2{}\fi
\expandafter\ifx\csname urllinklabel\endcsname\relax
  \def\urllinklabel{[LINK]}\fi
\expandafter\ifx\csname adsurllinklabel\endcsname\relax
  \def\adsurllinklabel{[ADS]}\fi



\bibitem[{{Bakos} {et~al.}(2006)}]{Bako06}
{Bakos}, G.~{\'A}., {Knutson}, H., {Pont}, F., {Moutou}, C., 
	{Charbonneau}, D., {Shporer}, A., {Bouchy}, F., {Everett}, M., 
	{Hergenrother}, C., {Latham}, D.~W., {Mayor}, M., {Mazeh}, T., 
	{Noyes}, R.~W., {Queloz}, D., {P{\'a}l}, A., {Udry}, S. 2006, ApJ, 650, 1160


\bibitem[{{Barnes} {et~al.}(1998)}]{barn98}
{Barnes}, J. R., {Collier Cameron}, A., {Unruh}, Y. C., {Donati}, J. F., {Hussain}, G. A. J., 1998, MNRAS, 299, 904

\bibitem[{{Barnes} {et~al.}(2007a)}]{2007MNRAS.379.1097B}
{Barnes}, J.~R., {Leigh}, C.~J., {Jones}, H.~R.~A., {Barman}, T.~S.,
  {Pinfield}, D.~J., {Collier Cameron}, A., \& {Jenkins}, J.~S. 2007a, MNRAS,
  379, 1097

\bibitem[{{Barnes} {et~al.}(2007b)}]{barn07b}
{Barnes}, J.~R., {Barman}, T. S., {Prato}, L., {Segransan}, D., {Jones}, H. R. A., {Leigh}, C.~J.,  {Collier Cameron}, A.,  {Pinfield}, D.~J., 2007b, MNRAS,
  382, 473


\bibitem[{{Barnes} {et~al.}(2008){Barnes}, {Barman}, {Jones}, {Leigh}, {Collier
  Cameron}, {Barber}, \& {Pinfield}}]{bar08}
{Barnes}, J.~R., {Barman}, T.~S., {Jones}, H.~R.~A., {Leigh}, C.~J., {Collier
  Cameron}, A., {Barber}, R.~J., \& {Pinfield}, D.~J. 2008, MNRAS, 390, 1258


\bibitem[{{Barnes} {et~al.}(2010){Barnes}, {Barman}, {Jones}, {Barber},
  {Hansen}, {Prato}, {Rice}, {Leigh}, {Collier Cameron}, \& {Pinfield}}]{bar10}
{Barnes}, J.~R., {Barman}, T.~S., {Jones}, H.~R.~A., {Barber}, R.~J., {Hansen},
  B.~M.~S., {Prato}, L., {Rice}, E.~L., {Leigh}, C.~J., {Collier Cameron}, A.,
  \& {Pinfield}, D.~J. 2010, MNRAS, 401, 445


\bibitem[{{Barrow}{ et~al.}(1982){Barrow},{Bhavsar},\&{Sonoda}}]{bar92}{Barrow}, J.~D., {Bhavsar}, S.~P. \& {Sonoda}, D.~H. 1982, MNRAS, 210, 19


\bibitem[{{Bean}{ et~al.}(2010)}]{Bean10}{Bean}, J.~L., {Miller-Ricci Kempton}, E., {Homeier}, D. 2010, Nature, 468, 669


\bibitem[{Brogi} {et~al.}(2012)]{Brog12}
{Brogi}, M., {Snellen}, I.\,A.\,G., {de Kok}, R.\,J., {Albrecht}, S., {Birkby},
  J., {de Mooij}, E.\,J.\,W. 2012,  Nature 486, 502

\bibitem[{{Burrows} \& {Sharp}(1999)}]{1999ApJ...512..843B}
{Burrows}, A. \& {Sharp}, C.~M. 1999, ApJ, 512, 843


\bibitem[{{Bouchy} {et~al.}(2005)}]{Bouc05}
{Bouchy}, F., {Udry}, S., {Mayor}, M., {Moutou}, C., 
	{Pont}, F., {Iribarne}, N., {da Silva}, R., {Ilovaisky}, S., 
	{Queloz}, D., {Santos}, N.~C., {S{\'e}gransan}, D., 
	{Zucker}, S. 2005, A\&A, 444, 15

\bibitem[{{Charbonneau} {et~al.}(1999){Charbonneau}, {Noyes}, {Korzennik},
  {Nisenson}, {Jha}, {Vogt}, \& {Kibrick}}]{1999ApJ...522L.145C}
{Charbonneau}, D., {Noyes}, R.~W., {Korzennik}, S.~G., {Nisenson}, P., {Jha},
  S., {Vogt}, S.~S., \& {Kibrick}, R.~I. 1999, ApJl, 522, L145

\bibitem[{{Charbonneau} {et~al.}(2002)}]{Char02}{Charbonneau}, D., {Brown}, T.~M., {Noyes}, R.~W., 
	{Gilliland}, R.~L. 2002, ApJ, 568, 377	

\bibitem[{{Charbonneau} {et~al.}(2008)}]{Char08}
{Charbonneau}, D., {Knutson}, H.~A., {Barman}, T., 
	{Allen}, L.~E., {Mayor}, M., {Megeath}, S.~T., {Queloz}, D., 
	{Udry}, S. 2008, ApJ, 686, 1341
	
\bibitem[{{Collier Cameron} {et~al.}(2000)}]{coli00}
{Collier Cameron}, A., {Horne}, K., {Penny}, A., \& {James}, D. 2000, Nature,
  402, 751

\bibitem[{{Collier Cameron} {et~al.}(2002){Collier Cameron}, {Horne}, {Penny},
  \& {Leigh}}]{2002MNRAS.330..187C}
{Collier Cameron}, A., {Horne}, K., {Penny}, A., \& {Leigh}, C. 2002, MNRAS,
  330, 187


\bibitem[{{Cooper} \& {Showman}(2006)}]{2006ApJ...649.1048C}
{Cooper}, C.~S. \& {Showman}, A.~P. 2006, ApJ, 649, 1048


\bibitem[{{Col{\'o}n} {et~al.}(2012)}]{Colo12}{Col{\'o}n}, K.~D., {Ford}, E.~B., {Redfield}, S., 
	{Fortney}, J.~J., {Shabram}, M., {Deeg}, H.~J., {Mahadevan}, S. 2012, MNRAS, 419, 2233
	
	
	\bibitem[{{Cubillos} {et~al.}(2011){Cubillos}, {Rojo}, \& {Fortney}}]{cub11}
{Cubillos}, P.~E., {Rojo}, P., \& {Fortney}, J.~J. 2011, A\&A, 529, A88

\bibitem[{{de Mooij} {et~al.}(2012)}]{deMo12}{de Mooij}, E.~J.~W., {Brogi}, M., {de Kok}, R.~J., 
	{Koppenhoefer}, J., {Nefs}, S.~V., {Snellen}, I.~A.~G., 
	{Greiner}, J., {Hanse}, J., {Heinsbroek}, R.~C., {Lee}, C.~H., 
	{van der Werf}, P.~P. 2012, A\&A, 538, 46
	
	
	\bibitem[{{Deming} {et~al.}(2005)}]{Demi05}
{Deming}, D., {Brown}, T.~M., {Charbonneau}, D., {Harrington}, J., 
	{Richardson}, L.~J. 2005, ApJ, 622, 1149

\bibitem[{{Donati} {et~al.}(1997)}]{dona97}
{Donati}, J.-F., {Semel}, M., {Carter}, B., {Rees}, D. E., {Collier Cameron}, A., 1997, MNRAS, 291, 658



\bibitem[{{D{\'e}sert} {et~al.}(2009){D{\'e}sert}, {Lecavelier des Etangs},
  {H{\'e}brard}, {Sing}, {Ehrenreich}, {Ferlet}, \&
  {Vidal-Madjar}}]{2009ApJ...699..478D}
{D{\'e}sert}, J.-M., {Lecavelier des Etangs}, A., {H{\'e}brard}, G., {Sing},
  D.~K., {Ehrenreich}, D., {Ferlet}, R., \& {Vidal-Madjar}, A. 2009, ApJ, 699,
  478
\bibitem[{{Endl} {et~al.}(2000)}]{Endl00}
{Endl}, M., {K{\"u}rster}, M., {Els}, S. 2000, A\&A, 362, 585



\bibitem[{{Grillmair} {et~al.}(2008){Grillmair}, {Burrows}, {Charbonneau},
  {Armus}, {Stauffer}, {Meadows}, {van Cleve}, {von Braun}, \&
  {Levine}}]{2008Natur.456..767G}
{Grillmair}, C.~J., {Burrows}, A., {Charbonneau}, D., {Armus}, L., {Stauffer},
  J., {Meadows}, V., {van Cleve}, J., {von Braun}, K., \& {Levine}, D. 2008,
  Nature, 456, 767


\bibitem[{{Gustafsson} {et~al.}(2008){Gustafsson}, {Edvardsson}, {Eriksson},
  {J{\o}rgensen}, {Nordlund}, \& {Plez}}]{2008A&A...486..951G}
{Gustafsson}, B., {Edvardsson}, B., {Eriksson}, K., {J{\o}rgensen}, U.~G.,
  {Nordlund}, {\AA}., \& {Plez}, B. 2008, A\&A, 486, 951


\bibitem[{{Hauschildt} {et~al.}(1997){Hauschildt}, {Baron}, \&
  {Allard}}]{1997ApJ...483..390H}
{Hauschildt}, P.~H., {Baron}, E., \& {Allard}, F. 1997, ApJ, 483, 390


\bibitem[{{Helling} {et~al.}(2008){Helling}, {Dehn}, {Woitke}, \&
  {Hauschildt}}]{2008ApJ...675L.105H}
{Helling}, C., {Dehn}, M., {Woitke}, P., \& {Hauschildt}, P.~H. 2008, ApJl,
  675, L105


\bibitem[{{Henry} {et~al.}(2000){Henry}, {Baliunas}, {Donahue}, {Fekel}, \&
  {Soon}}]{2000ApJ...531..415H}
{Henry}, G.~W., {Baliunas}, S.~L., {Donahue}, R.~A., {Fekel}, F.~C., \& {Soon},
  W. 2000, ApJ, 531, 415


\bibitem[{{Kaeufl} {et~al.}(2004){Kaeufl}, {Ballester}, {Biereichel},
  {Delabre}, {Donaldson}, {Dorn}, {Fedrigo}, {Finger}, {Fischer}, {Franza},
  {Gojak}, {Huster}, {Jung}, {Lizon}, {Mehrgan}, {Meyer}, {Moorwood}, {Pirard},
  {Paufique}, {Pozna}, {Siebenmorgen}, {Silber}, {Stegmeier}, \&
  {Wegerer}}]{Kaeu06}
{Kaeufl}, H.-U., {Ballester}, P., {Biereichel}, P., {Delabre}, B., {Donaldson},
  R., {Dorn}, R., {Fedrigo}, E., {Finger} et~al. 2004, SPIE, 5492, 1218

\bibitem[{{Knutson} {et~al.}(2007)}]{Knut07}
{Knutson}, H.~A., {Charbonneau}, D., {Allen}, L.~E., 
	{Fortney}, J.~J., {Agol}, E., {Cowan}, N.~B., {Showman}, A.~P., 
	{Cooper}, C.~S., {Megeath}, S.~T. 2007, Nature, 447, 183

\bibitem[{{Knutson} {et~al.}(2012)}]{Knut12}
{Knutson}, H.~A., {Lewis}, N., {Fortney}, J.~J., {Burrows}, A., 
	{Showman}, A.~P., {Cowan}, N.~B., {Agol}, E., {Aigrain}, S., 
	{Charbonneau}, D., {Deming}, D., {D{\'e}sert}, J.-M., 
	{Henry}, G.~W., {Langton}, J., {Laughlin}, G. 2012, ApJ, 754, 22

\bibitem[{{K\"urster} {et~al.}(1997){Kuerster}, {Schmitt}, {Cutispoto}, \&
  {Dennerl}}]{1997A&A...320..831K}
{K\"urster}, M., {Schmitt}, J.~H.~M.~M., {Cutispoto}, G., \& {Dennerl}, K. 1997,
  A\&A, 320, 831
  
\bibitem[{{Langford} {et~al.}(2011)}]{Lang11}
{Langford}, S.~V., {Wyithe}, J.~S.~B., {Turner}, E.~L., 
	{Jenkins}, E.~B., {Narita}, N., {Liu}, X., {Suto}, Y., 
	{Yamada}, T. 2011, MNRAS, 415, 673

\bibitem[{{Lecavelier Des Etangs} {et~al.}(2010)}]{Leca10}{Lecavelier Des Etangs}, A., {Ehrenreich}, D., {Vidal-Madjar}, A., 
	{Ballester}, G.~E., {D{\'e}sert}, J.-M., {Ferlet}, R., 
	{H{\'e}brard}, G., {Sing}, D.~K., {Tchakoumegni}, K.-O., 
	{Udry}, S. 2010, A\&A, 514, 72

\bibitem[{{Leigh} {et~al.}(2003){Leigh}, {Collier Cameron}, {Horne}, {Penny},
  \& {James}}]{2003MNRAS.344.1271L}
{Leigh}, C., {Collier Cameron}, A., {Horne}, K., {Penny}, A., \& {James}, D.
  2003, MNRAS, 344, 1271


\bibitem[{{L{\'o}pez-Morales} \& {Seager}(2007)}]{2007ApJ...667L.191L}
{L{\'o}pez-Morales}, M. \& {Seager}, S. 2007, ApJl, 667, L191



\bibitem[{{L{\'o}pez-Morales} {et~al.}(2010)}]{Lope10}{L{\'o}pez-Morales}, M., {Coughlin}, J.~L., {Sing}, D.~K., 
	{Burrows}, A., {Apai}, D., {Rogers}, J.~C., {Spiegel}, D.~S., 
	{Adams}, E.~R. 2010, ApJ, 716, 36
	
	
	
\bibitem[{{Marley} {et al.}(1999)}]{Marl99}
{Marley}, M.\,S., {Gelino}, C., {Stephens}, D., {Lunine}, J.\,I., {Freedman},
  R. 1999,  ApJ 513, 879
  
\bibitem[{{McLean} {et~al.}(1998)}]{Mcle98}{McLean}, I. S., {Becklin}, E. E., {Bendiksen}, O. et~al. 1998, Proc. SPIE, 3354, 566
 
 \bibitem[{{Pont} {et al.}(2008)}]{Pont08}
 {Pont}, F., {Knutson}, H., {Gilliland}, R.~L., {Moutou}, C., 
	{Charbonneau}, D. 2008, MNRAS, 385, 109
 
 
 \bibitem[{{Press} {et al.}(1992){Press}, {Teukolsky}, {Vetterling}, \& {Flannery}}]{pre92} {Press}, W.~H., {Teukolsky}, S.~A., {Vetterling} W.~T., \&
 {Flannery} B.~P., 1992,   Numerical recipes in C. The art of scientific
 computing, Cambridge University Press

\bibitem[{{Redfield} {et~al.}(2008)}]{Redf08}{Redfield}, S., {Endl}, M., {Cochran}, W.~D., {Koesterke}, L. 2008, ApJ, 673, 87


\bibitem[{{Rodler} {et~al.}(2008){Rodler}, {K{\"u}rster}, \&
  {Henning}}]{2008A&A...485..859R}
{Rodler}, F., {K{\"u}rster}, M., \& {Henning}, T. 2008, A\&A, 485, 859


\bibitem[{{Rodler} {et~al.}(2010){Rodler}, {K{\"u}rster}, \& {Henning}}]{rod10}
{Rodler}, F., {K{\"u}rster}, M., \& {Henning}, T. 2010, A\&A, 514, A23

\bibitem[{{Rodler} {et~al.}(2012)}]{Rodl12b}
{Rodler}, F., {Lopez-Morales}, M., {Ribas}, I. 2012,  ApJ, 753, L25



\bibitem[{{Seager} {et~al.}(1998)}]{1998ApJ...502L.157S}{Seager}, S., \& {Sasselov}, D.~D. 1998, ApJ, 502, 157



\bibitem[{{Sharp} \& {Burrows}(2007)}]{2007ApJS..168..140S}
{Sharp}, C.~M. \& {Burrows}, A. 2007, ApJs, 168, 140


\bibitem[{{Sing} \& {L{\'o}pez-Morales}(2009)}]{Sing09}{Sing}, D.~K., {L{\'o}pez-Morales}, M. 2009, A\&A, 493, 31


\bibitem[{{Sing} {et~al.}(2011)}]{Sing11}{Sing}, D.~K., {Pont}, F., {Aigrain}, S., {Charbonneau}, D., 
	{D{\'e}sert}, J.-M., {Gibson}, N., {Gilliland}, R., 
	{Hayek}, W. et al. 2011, MNRAS, 416, 1443
	
\bibitem[{{Snellen} {et~al.}(2008)}]{Snel08}{Snellen}, I.~A.~G., {Albrecht}, S., {de Mooij}, E.~J.~W., 
	{Le Poole}, R.~S. 2008, A\&A, 487, 357

\bibitem[{{Snellen} {et~al.}(2010){Snellen}, {de Kok}, {de Mooij}, \&
  {Albrecht}}]{Sne10}
{Snellen}, I.~A.~G., {de Kok}, R.~J., {de Mooij}, E.~J.~W., \& {Albrecht}, S.
  2010, Nature, 465, 1049

\bibitem[{{Sudarsky} {et~al.}(2000)}]{Suda00}
{Sudarsky}, D., {Burrows}, A., {Pinto}, P. 2000,  ApJ, 538, 885

\bibitem[{{Sudarsky} {et~al.}(2003)}]{Suda03}
{Sudarsky}, D., {Burrows}, A., {Hubeny}, I.: 2003,  ApJ, 588, 1121

\bibitem[{{Swain} {et~al.}(2009){Swain}, {Vasisht}, {Tinetti}, {Bouwman},
  {Chen}, {Yung}, {Deming}, \& {Deroo}}]{2009ApJ...690L.114S}
{Swain}, M.~R., {Vasisht}, G., {Tinetti}, G., {Bouwman}, J., {Chen}, P.,
  {Yung}, Y., {Deming}, D., \& {Deroo}, P. 2009, ApJl, 690, L114


\bibitem[{{Tinetti} {et~al.}(2007)}]{07}
{Tinetti}, G., {Vidal-Madjar}, A., {Liang}, M.-C., 
	{Beaulieu}, J.-P., {Yung}, Y., {Carey}, S., {Barber}, R.~J., 
	{Tennyson}, J., {Ribas}, I., {Allard}, N., {Ballester}, G.~E., 
	{Sing}, D.~K., {Selsis}, F. 2007, Nature, 448, 169





\bibitem[{{van Belle} \& {von Braun}(2009)}]{2009ApJ...694.1085V}
{van Belle}, G.~T. \& {von Braun}, K. 2009, ApJ, 694, 1085

\bibitem[{{Vidal-Madjar} {et~al.}(2003)}]{Vida03}{Vidal-Madjar}, A., {Lecavelier des Etangs}, A., {D{\'e}sert}, J.-M., 
	{Ballester}, G.~E., {Ferlet}, R., {H{\'e}brard}, G., 
	{Mayor}, M. 2003, Nature, 422, 143

\bibitem[{{Vidal-Madjar} {et~al.}(2004)}]{Vida04}{Vidal-Madjar}, A., {D{\'e}sert}, J.-M., {Lecavelier des Etangs}, A., 
	{H{\'e}brard}, G., {Ballester}, G.~E., {Ehrenreich}, D., 
	{Ferlet}, R., {McConnell}, J.~C., {Mayor}, M., {Parkinson}, C.~D. 2004, ApJ, 604, 69	

\bibitem[{{Waldmann} {et~al.}(2012)}]{Wald12}{Waldmann}, I.~P., {Tinetti}, G., {Drossart}, P., {Swain}, M.~R., 
	{Deroo}, P., {Griffith}, C.~A. 2012, ApJ, 744, 35
	
	
\bibitem[{{Wiedemann} {et~al.}(2001){Wiedemann}, {Deming}, \&
  {Bjoraker}}]{2001ApJ...546.1068W}
{Wiedemann}, G., {Deming}, D., \& {Bjoraker}, G. 2001, ApJ, 546, 1068


\bibitem[{{Witte} {et~al.}(2011){Witte}, {Helling}, {Barman}, {Heidrich}, \&
  {Hauschildt}}]{2011A&A...529A..44W}
{Witte}, S., {Helling}, C., {Barman}, T., {Heidrich}, N., \& {Hauschildt},
  P.~H. 2011, A\&A, 529, A44

\end{thebibliography}
\end{document}